\documentclass[10pt,final]{article}

\usepackage{graphicx}
\usepackage{dcolumn}
\usepackage{bm}
\usepackage{amssymb}
\usepackage{amsmath}
\usepackage[square,numbers]{natbib}
\usepackage{authblk}
\usepackage{braket}
\usepackage{amsopn}

\DeclareMathOperator{\Tr}{Tr}
\DeclareMathOperator{\spa}{Span}

\title{$\mathcal{PT}$-symmetric interference transistor}

\author[1,2,*]{Alexander A. Gorbatsevich}
\author[2]{Gennadiy Ya. Krasnikov}
\author[1,2,3]{Nikolay M. Shubin}
\affil[1]{P.N. Lebedev Physical Institute of the Russian Academy of Sciences, Division of solid state physics, Moscow, 119991, Russia.}
\affil[2]{JSC Molecular Electronics Research Institute, Zelenograd, Moscow, 124460, Russia.}
\affil[3]{National Research University of Electronic Technology, Department of quantum physics and nanoelectronics, Zelenograd, Moscow, 124498, Russia.}

\affil[*]{aagor137@mail.ru}

\begin{document}

\maketitle

\begin{abstract}
We present a model of the molecular transistor, operation of which is based on the interplay between two physical mechanisms, peculiar to open quantum systems that act in concert: $\mathcal{PT}$-symmetry breaking corresponding to coalescence of resonances at the exceptional point of the molecule, connected to the leads, and Fano-Feshbach antiresonance. This switching mechanism can be realised in particular in a special class of molecules with degenerate energy levels, i.e. diradicals, which possess mirror symmetry. At zero gate voltage infinitesimally small interaction of the molecule with the leads breaks the $\mathcal{PT}$-symmetry of the system that, however, can be restored by application of the gate voltage preserving the mirror symmetry. $\mathcal{PT}$-symmetry broken state at zero gate voltage with minimal transmission corresponds to the `off' state while the $\mathcal{PT}$-symmetric state at non-zero gate voltage with maximum transmission -- to the `on' state. At zero gate voltage energy of the antiresonance coincides with exceptional point but the transmission variation mainly takes place due to the coalescence of resonances at the exceptional point. We construct a model of an all-electrical molecular switch based on such transistors acting as a conventional CMOS inverter and show that essentially lower power consumption and switching energy can be achieved, compared to the CMOS analogues.
\end{abstract}

\thispagestyle{empty}

\section{Introduction}
Implementation of molecules in integrated circuits (IC) offers great advantages due to extreme miniaturization and perfect reproducibility.\cite{bib:Mol1,bib:Mol2,bib:SUPERBOOK} But despite long-term and intensive efforts since its origin in the early 70s,\cite{bib:Aviram} molecular electronics (ME) has not yet presented any experimentally realized candidate to replace the silicon transistor as a `wheel-horse' of the modern IC industry. High expectations were held and are still in place with graphene \cite{bib:Geim} and post graphene organic Dirac materials.\cite{bib:PostGraph} During past period ME mainly concentrated on the attempts to reproduce typical elements of silicon electronics.\cite{bib:Joach2005,bib:Schwierz2010,bib:Zhao2016,bib:CNTHan2,bib:Cao1369,bib:Iannaccone2018} In the case of graphene and related materials this approach has been based on the efforts to develop band opening methods,\cite{bib:Zhang2009} which, however, haven't resulted yet in a new IC technology either. On the other hand, due to complex geometry and topology of molecular structures one could expect that the devices with working principles, different from the ordinary field-effect and bipolar transistors, could be designed.

Energy spectrum of a molecule manifests itself in transport phenomena by means of resonances. If the molecule possesses different carrier paths, destructive interference can result in formation of asymmetric Fano-Feshbach resonance,\cite{bib:FanoRMP} which combines a resonance (transmission peak) and an antiresonance (transmission dip) nearby. Existence of the interference effect in transport through molecules, which is intensively discussed  in the literature,\cite{bib:Lambert2015,bib:Markussen2010,bib:Peder2014,bib:Tsuji2014,bib:solomon2015Book} is now well established experimentally.\cite{bib:Fracasso2011,bib:Guedon2012,bib:Vazquez2012,bib:Aradhya2012} In Ref.~\cite{bib:Mazumdar2007} quantum interference transistor (QIT) was described with the `off' state corresponding to perfect interference destruction of both transmission and current. One of the main challenges in CMOS electronics is reduction of the operating voltage that doesn't follow Moore's law (ITRS 2.0). In Ref.~\cite{bib:MolTransSciRep} it was argued that the interference control of the carrier transport over different paths can substantially reduce the operating gate voltage, because the suppression of the transmission function can be achieved at lower gate voltage compared with the one required to move the transmission function peak away from the distribution function window. However, antiresonances, which arise from the destructive quantum interference (DQI), are determined by the topology of the structure that includes different interfering carrier paths. Hence, variation of the on-site potential and/or intersite hopping can only shift the antiresonance in energy rather than destroy it, because interfering paths are retained under such variations. The voltage required to shift an antiresonance away from the operating energy region is determined by the carrier distribution in the leads on a scale no less than $kT$ and, hence, is not small. Therefore, the proposed control of the transmission resonance by low voltages should rely on a mechanism more complex than multipath interference solely. For a logical gate to operate, its constituting elements (transistors) should undergo transitions between the `off' and the `on' states, with the latter state being even more important than the former one as it provides switching of the successive gate. The `on'/`off' ratio for the transistor conductance should be as high as possible to provide a reliable gate operation. However, this requirement is scarcely achievable in quantum interference transistors operating near the antiresonance because of the low transmission away from the antiresonance.\cite{bib:GorShUFN} Hence, a quantum  transistor is required, which possesses a combination of antiresonance and nearby resonance that is responsible for high conductance in the `on' state. 

In this paper we show that, indeed, the transmission probability of a special class of molecules can be controlled in a wide range by applying small gate voltages due to the interplay of two physical mechanisms: $\mathcal{PT}$-symmetry breaking, accompanied by the collapse of resonances\cite{bib:Gor} at the exceptional point (EP) of the molecule connected to the leads,\cite{bib:GorShAop,bib:GorShPRB} and the shift of Fano-Feshbach resonance to the EP point. This special class consists of the molecules with degenerate energy levels, e.g. diradicals\cite{bib:DirRev,bib:ZeroDirad,bib:Nozaki2017a} (but not restricted to), which possess mirror symmetry.

\section{Phenomenological model}
Consider an open quantum system comprised of a molecule and contacts that possesses EP in a sense of Ref.~\cite{bib:GorShPRB}. At this EP two unity resonances coalesce and cancel each other making the transition to the `off' state very sharp. An open quantum system should be spatially symmetric in order to possess EP. To take advantage of both DQI and coalescence of resonances at the EP one should consider a system with two resonances and one antiresonance. The transmission coefficient of an arbitrary two-terminal quantum system can be written in the compact form:~\cite{bib:GorShAop,bib:GorShPRB}
\begin{equation}
T(\omega)=\frac{|P(\omega)|^{2}}{|P(\omega)|^{2}+|Q(\omega)|^{2}}.
\label{eq:TPQ}
\end{equation}
Here $P(\omega)$ and $Q(\omega)$ are some functions of an energy $\omega$. Real zeroes of function $P(\omega)$ correspond to transmission nodes (antiresonances), while real zeroes of function $Q(\omega)$ determine exact positions of perfect (unity) resonances on the energy axis.\cite{bib:GorShAop,bib:GorShPRB} In the vicinity of the resonances and antiresonance $P(\omega)$ and $Q(\omega)$ can be expressed as:\cite{bib:GorShPRB}
\begin{equation}
\begin{split}
P(\omega)&=2\Gamma B\left(\omega-\varepsilon_{0}\right)D_{P},\\
Q(\omega)&=\left(\omega-\varepsilon_{1}^{+}\right)\left(\omega-\varepsilon_{1}^{-}\right)D_{Q},
\end{split}
\label{eq:PQrar}
\end{equation}
where $\varepsilon_{0}$ and $\varepsilon_{1}^{\pm}$ determine exact position of the transmission antiresonance and resonances, correspondingly, $\Gamma$ is the imaginary part of the contact self-energy describing interaction of a molecule with the leads\cite{bib:DattaBook1997,bib:Ryndyk} and $B$ is some positive dimensionless coefficient. Factors $D_{P}$ and $D_{Q}$ take into account the contributions from the remote energy levels and can be estimated as $D_{P}\sim D_{Q}\sim\Delta^{N-2}$, where $\Delta$ is an average distance between the remote energy levels and $N$ is the dimension of the molecular orbital Hilbert space. Phenomenologically, functions $P(\omega)$ and $Q(\omega)$ are defined up to an arbitrary common factor, hence, we can redefine the parameter $B\mapsto BD_{P}/D_{Q}$ and replace three phenomenological parameters $B$, $D_{P}$ and $D_{Q}$ by just $B$. Further we will use $B$ as such generalized parameter.

Consider a model that possesses degenerate antiresonance and resonance levels in the symmetric phase, which can be distorted by an external perturbation described by parameter $\delta$. Energies of the antiresonance $\varepsilon_{0}$ and resonances $\varepsilon_{1}^{\pm}$ can be expressed as:
\begin{equation}
\begin{split}
\varepsilon_{0}&=x_{0}\delta,\\
\varepsilon_{1}^{\pm}&=x_{1}\delta\pm\sqrt{y^{2}\delta^{2}-z^{2}\Gamma^{2}}.
\end{split}
\label{eq:rarEnerg}
\end{equation}
Here $x_{0,1}$, $y$ and $z$ are some dimensionless parameters depending on the structure of a particular system. Terms in Eq.~(\ref{eq:rarEnerg}), which are linear in $\delta$, describe the shift of the (anti)resonance positions due to the external perturbation and non-analytical term (square root) in the expression for $\varepsilon_{1}^{\pm}$ describes the coalescence of resonances phenomenon. Energy of the degenerate state (at $\delta=0$) is set to the energy origin. If the external perturbation $\delta$ is high enough ($\delta>zy^{-1}\Gamma$), then the transmission has two unity peaks at $\omega=\varepsilon_{1}^{\pm}$, which coalesce at $\delta=zy^{-1}\Gamma$. The poorest transmission profile (i.e. the `off' state) corresponds to $\delta=0$. From Eqs.~(\ref{eq:TPQ}-\ref{eq:rarEnerg}) one can see, that in this case there are two peaks at $\omega=\pm z\Gamma$ with
\begin{equation}
T_{peak}\left(\omega=\pm z\Gamma;\delta=0\right)=\frac{B^{2}}{B^{2}+z^{2}},
\label{eq:maxTraPhenom}
\end{equation}
separated by a zero dip at $\omega=0$.

\section{Microscopic model}
The microscopic model of the system, transmission coefficient of which possesses the phenomenological properties described above, is as follows. There are two degenerate states $\ket{1}$ and $\ket{2}$ with the same energy $\varepsilon$. This system is attached symmetrically to two leads (left and right) in such a way that the mirror symmetry operation $\sigma_{LR}$, which maps the left lead into the right one and vice versa, is also an element of the symmetry group $G$ of the bare Hamiltonian of the system, i.e. $\sigma_{LR}\in G$. Due to the degeneracy, there must be an irreducible representation of the symmetry group $G$ acting on the subspace $\mathcal{H}_{12}=\spa{(\ket{1},\ket{2})}$ of the total Hilbert space of states of the isolated system. Let us choose the basis in $\mathcal{H}_{12}$ as the basis of a symmetric $\ket{s}$ and an anti-symmetric $\ket{a}$ states, which are the eigenstates of the reflection operator $\sigma_{LR}$: $\sigma_{LR}\ket{s}=\ket{s}$ and $\sigma_{LR}\ket{a}=-\ket{a}$. These states conserve their symmetry with introduction of the perturbation, which is invariant under $\sigma_{LR}$. The tunnelling matrix elements between the leads and the symmetric state are of the same sign, whereas, the tunnelling matrix elements between the leads and the anti-symmetric state are of opposite signs (see Fig.~\ref{fig10}a). Therefore in this basis couplings to the leads can be written as
\begin{equation}
\begin{split}
\mathbf{u}_{L}&=\sqrt{\Gamma}\begin{pmatrix}
												\gamma_{s}\\
												\gamma_{a}
												\end{pmatrix},\\
\mathbf{u}_{R}&=\sqrt{\Gamma}\begin{pmatrix}
												\gamma_{s}\\
												-\gamma_{a}
												\end{pmatrix}.
\end{split}
\label{eq:saCoupl}
\end{equation}
Here $\Gamma$ governs the coupling strength and positive dimensionless parameters $0\leq\gamma_{s,a}\leq1$ describe relative couplings of symmetric and anti-symmetric states to the leads. Parameters $\gamma_{s,a}$ can be calculated, for example, as projections of the vector $\Gamma^{-1/2}\mathbf{u}_{L,R}^{site}$ onto $\ket{s}$ or $\ket{a}$ respectively, where $\mathbf{u}_{L,R}^{site}$ describes the coupling to the leads in the site (atomic orbitals) basis. If each lead is attached to only one site, then $\gamma_{s,a}$ is just a contribution of the state localized in the connection site to the symmetric or anti-symmetric state correspondingly (see Supplementary materials).

Application of the gate voltage introduces external perturbation that lowers the symmetry of the system, resulting in removal of the degeneracy. Suppose that the external perturbation lowers the symmetry of the system from the group $G$ to its some non-trivial subgroup $H\subset G$, such that $\sigma_{LR}\in H$. This perturbation introduces detuning of the energy of symmetric and anti-symmetric states: $\varepsilon_{s,a}(\delta)=\varepsilon+k_{s,a}\delta$ with $\delta>0$ and dimensionless parameters $-1\leq k_{s,a}\leq1$ accounting for the different influence of the perturbation on the energies of symmetric and anti-symmetric states (see Fig.~\ref{fig10}a). So, the bare Hamiltonian of the system becomes following:
\begin{equation}
\hat H_{0}=\begin{pmatrix}
				\varepsilon+k_{s}\delta & 0\\
				0 & \varepsilon+k_{a}\delta
			  \end{pmatrix}
\label{eq:H0Degen}
\end{equation}
Parameters $k_{s,a}$ can be estimated, for instance, from the perturbation theory (see Supplementary materials for details); note that $k_{a}\neq k_{s}$ as the considered perturbation removes the degeneracy. Assume that couplings~(\ref{eq:saCoupl}) are not affected by this perturbation. In fact, $\Gamma$ and $\gamma_{s,a}$ are some smooth functions of the perturbation strength, i.e. $\delta$. However, taking this into account does not change the qualitative picture described below.

\subsection{Transport properties}
Tunnelling transmission coefficient through the states $\ket{s}$ and $\ket{a}$ (neglecting the contribution from remote states to the transport process) can be calculated by the standard formula:\cite{bib:DattaBook1997}
\begin{equation}
T=4\Tr{\left(\hat \Gamma_{R}\hat G^{r}\hat \Gamma_{L}\hat G^{a}\right)},
\label{eq:STDTran}
\end{equation}
where $\hat G^{r,a}$ is the retarded/advanced Green function and $\hat \Gamma_{L,R}=\mathbf{u}_{L,R}\mathbf{u}_{L,R}^{\dag}$\cite{bib:HeffZelv} is the coupling matrix (imaginary part of corresponding contact self-energy) to the left or to the right lead. Here $\mathbf{u}_{L,R}$ are vectors, describing couplings of the states of the isolated system to the left/right lead. Traditional approach within the wide-band limit (neglecting real parts of the contact self-energy) leads to the following expression for the transmission:\cite{bib:GorShPRB}
\begin{equation}
T=\frac{4\left|\det{\left(\omega\hat I-\hat H_{0}\right)}\right|^{2}\left|\mathbf{u}_{R}^{\dag}\left(\omega\hat I-\hat H_{0}\right)^{-1}\mathbf{u}_{L}\right|^{2}}{\left|\det{\left(\omega\hat I-\hat H_{eff}\right)}\right|^{2}},
\label{eq:STDTranHeff}
\end{equation}
where $\hat H_{eff}=\hat H_{0}-i\mathbf{u}_{L}\mathbf{u}_{L}^{\dag}-\mathbf{u}_{R}\mathbf{u}_{R}^{\dag}$ is the Feshbach effective Hamiltonian. Following general formalism form Ref.~\cite{bib:GorShPRB}, one can show that transmission~(\ref{eq:STDTranHeff}) can be written in the form~(\ref{eq:TPQ}). For our microscopic model this can be easily checked using Eqs.~(\ref{eq:saCoupl}) and~(\ref{eq:H0Degen}). Indeed, the following identity holds true:
\begin{multline}
\left|Q\right|^{2}=\left|\det{\left(\omega\hat I-\hat H_{0}-i\mathbf{u}_{L}\mathbf{u}_{L}^{\dag}+\mathbf{u}_{R}\mathbf{u}_{R}^{\dag}\right)}\right|^{2}=\left|\det{\left(\omega\hat I-\hat H_{0}+i\mathbf{u}_{L}\mathbf{u}_{L}^{\dag}+\mathbf{u}_{R}\mathbf{u}_{R}^{\dag}\right)}\right|^{2}\\
-4\left|\det{\left(\omega\hat I-\hat H_{0}\right)}\right|^{2}\left|\mathbf{u}_{R}^{\dag}\left(\omega\hat I-\hat H_{0}\right)^{-1}\mathbf{u}_{L}\right|^{2}=\left[4\Gamma^2\gamma_{a}^{2}\gamma_{s}^{2}+\left(\omega-\varepsilon-k_{s}\delta\right)\left(\omega-\varepsilon-k_{a}\delta\right)\right]^{2}.
\label{eq:Calcs1}
\end{multline}
Hence, from Eq.~(\ref{eq:Calcs1}) we see that within the wide-band limit, the transmission of our system can be written in the form~(\ref{eq:TPQ}) with
\begin{equation}
\begin{split}
P\left(\omega\right)&=2\det{\left(\omega\hat I-\hat H_{0}\right)}\times\mathbf{u}_{L}^{\dag}\left(\omega\hat I-\hat H_{0}\right)^{-1}\mathbf{u}_{R},\\
Q\left(\omega\right)&=\det{\left(\omega\hat I-\hat H_{aux}\right)}.
\end{split}
\label{eq:PQDegen}
\end{equation}
Here
\begin{equation}
\hat H_{aux}=\hat H_{0}+i\mathbf{u}_{L}\mathbf{u}_{L}^{\dag}-i\mathbf{u}_{R}\mathbf{u}_{R}^{\dag}=\begin{pmatrix}
				\varepsilon+k_{s}\delta & 2i\Gamma\gamma_{s}\gamma_{a}\\
				2i\Gamma\gamma_{s}\gamma_{a} & \varepsilon+k_{a}\delta
			  \end{pmatrix}
\label{eq:HauxDegen}
\end{equation}
is the non-Hermitian auxiliary Hamiltonian with its real eigenvalues corresponding to energies of perfect transmission,\cite{bib:GorShPRB} and $\hat I$ is the $2\times2$ identity matrix. Hamiltonian $\hat H_{aux}$ is $\mathcal{PT}$-symmetric, where $\mathcal{P}=\sigma_{LR}$ denotes to the mirror reflection and $\mathcal{T}$ is the time reversal operator (complex conjugation). Indeed, one can easily check that operator $\mathcal{PT}\hat H_{aux}\mathcal{PT}$ acts on any vector $\mathbf{v}\in\mathbb{C}^{2}$ in the same way as operator $\hat H_{aux}$. Thus, $\hat H_{aux}$ is $\mathcal{PT}$-symmetric.\cite{bib:Bender2007} Therefore, it can possess real eigenvalues, which correspond to perfect transmission peaks, and for certain parameters they can coalesce and the $\mathcal{PT}$-symmetry of the Hamiltonian $\hat H_{aux}$ will be broken, leading to coalescence of perfect resonances into one peak with amplitude lower than $1$. Moreover, such resonance coalescence is accompanied by symmetry breaking of electron occupation at the energy corresponding to the transmission peak (see Appendix~\ref{ApA} for details).

\begin{figure}
\centering
\includegraphics[width=0.95\linewidth]{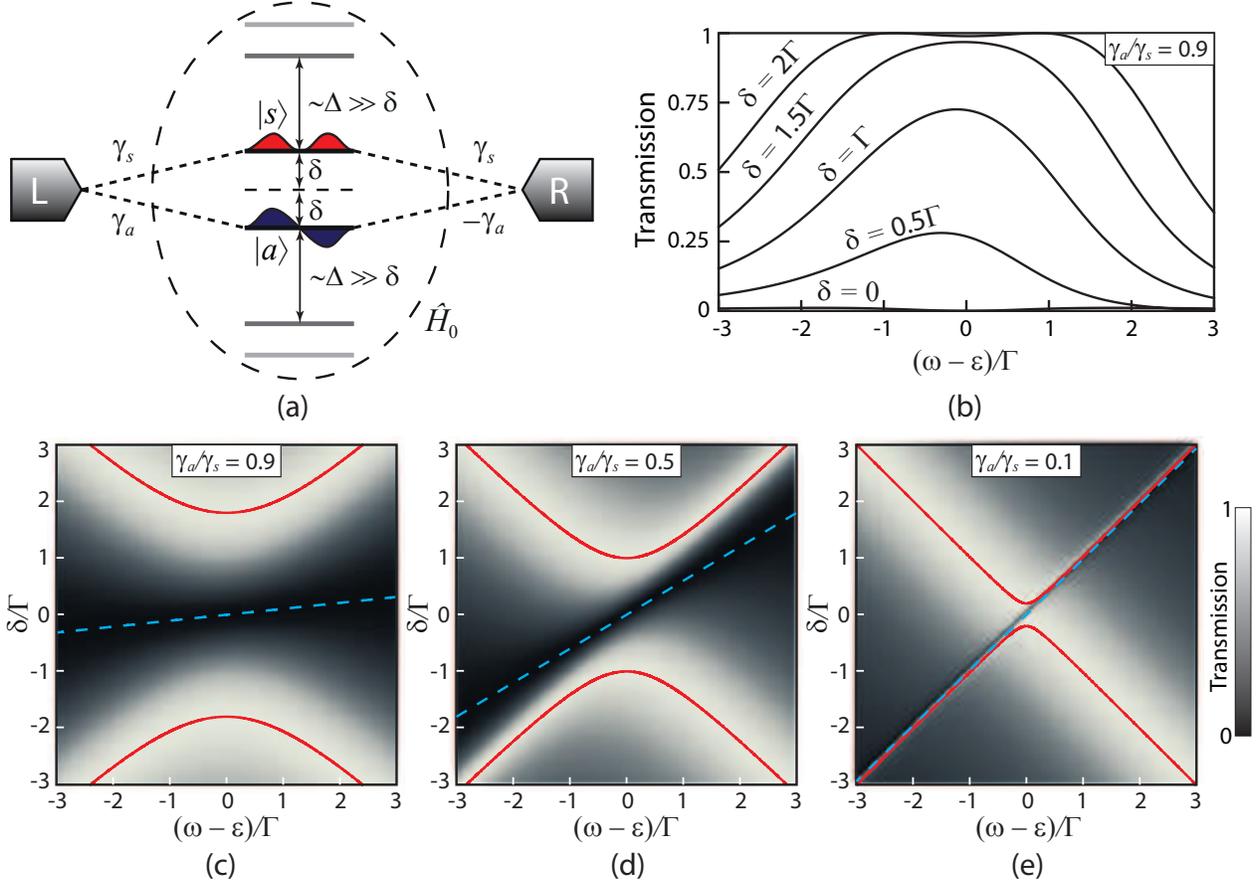}
\caption{Microscopic model and its transmission coefficient. (a) Schematic view of the microscopic model of the molecular system depicting symmetric $\ket{s}$ and anti-symmetric $\ket{a}$ states connected to the leads by couplings~(\ref{eq:saCoupl}). $\Gamma$ is set as energy unit and $k_{a}=-k_{s}=1$ for convenience. Evolution of the transmission coefficient profile with variation of $\delta$ ($\gamma_{s}$ is set to $1$ and $k_{a}=-k_{s}=1$) for (b) some discrete values of $\delta$ for $\gamma_{a}=0.9$, and (c-e) in the form of density plots for (c) $\gamma_{a}=1$, (d) $\gamma_{a}=0.5$ and (e) $\gamma_{a}=0.1$. Red solid lines indicate the position of perfect resonances and dashed cyan -- zeros of the transmission.}
\label{fig10}
\end{figure}

Using Eqs.~(\ref{eq:saCoupl}-\ref{eq:HauxDegen}), the transmission coefficient can be written as:
\begin{equation}
T\left(\omega\right)=\frac{4\Gamma^{2}\left[\left(k_{a}\gamma_{s}^{2}-k_{s}\gamma_{a}^{2}\right)\delta-\left(\gamma_{s}^{2}-\gamma_{a}^{2}\right)\left(\omega-\varepsilon\right)\right]^{2}}{4\Gamma^{2}\left[\left(k_{a}\gamma_{s}^{2}-k_{s}\gamma_{a}^{2}\right)\delta-\left(\gamma_{s}^{2}-\gamma_{a}^{2}\right)\left(\omega-\varepsilon\right)\right]^{2}+\left\{\left[\omega-\varepsilon-\frac12\delta\left(k_{a}+k_{s}\right)\right]^{2}+4\gamma_{s}^{2}\gamma_{a}^{2}\Gamma^{2}-\delta^{2}\frac{\left(k_{a}-k_{s}\right)^{2}}{4}\right\}^{2}}.
\label{eq:TraDegen}
\end{equation}
From this formula one can see that for a sufficiently large detuning $\delta$ there are two unity peaks of transmission (zeros of $Q$, i.e. eigenvalues of $\hat H_{aux}$) at $\omega=\varepsilon+\frac12\delta(k_{a}+k_{s})\pm\sqrt{\frac14\delta^{2}(k_{a}-k_{s})^{2}-4\gamma_{s}^{2}\gamma_{a}^{2}\Gamma^{2}}$. Decreasing the detuning one can achieve the coalescence of resonances at $\delta=4\gamma_{s}\gamma_{a}\Gamma|k_{a}-k_{s}|^{-1}$, which correspond to the EP of $\hat H_{aux}$. Further decreasing $\delta$ results in further lowering of the transmission coefficient peak. There is also a zero-valued antiresonance (zero of $P$) at $\omega=\varepsilon+\delta(k_{a}\gamma_{s}^{2}-k_{s}\gamma_{a}^{2})/(\gamma_{s}^{2}-\gamma_{a}^{2})$, which additionally lowers the transmission with decreasing $\delta$. One can see that, moving the energy origin to $\varepsilon$, Eq.~(\ref{eq:TraDegen}) takes the phenomenological form described by Eqs.~(\ref{eq:TPQ}-\ref{eq:rarEnerg}) with the following phenomenological parameters:
\begin{equation}
B=\left|\gamma_{s}^{2}-\gamma_{a}^{2}\right|\quad x_{0}=\frac{k_{a}\gamma_{s}^{2}-k_{s}\gamma_{a}^{2}}{\gamma_{s}^{2}-\gamma_{a}^{2}},\quad x_{1}=\frac{k_{a}+k_{s}}{2},\quad y=\frac{\left|k_{a}-k_{s}\right|}{2},\quad z=2\gamma_{s}\gamma_{a}.
\label{eq:PhenParam}
\end{equation}
Thus, according to Eq.~(\ref{eq:maxTraPhenom}) one can see that the poorest transmission peaks (at $\delta=0$) are
\begin{equation}
T_{peak}\left(\omega=\varepsilon\pm2\gamma_{s}\gamma_{a}\Gamma;\delta=0\right)=\frac{\left(\gamma_{s}^{2}-\gamma_{a}^{2}\right)^{2}}{\left(\gamma_{s}^{2}+\gamma_{a}^{2}\right)^{2}}.
\label{eq:maxTra}
\end{equation}
From~(\ref{eq:PhenParam}) we see that there is a limiting case $\gamma_{s}/\gamma_{a}\rightarrow1$ that results in $B\rightarrow0$ and $x_{0}\rightarrow\infty$, while the product $Bx_{0}\rightarrow\gamma_{s,a}^{2}(k_{a}-k_{s})$ remains finite. In this case complete opaqueness, i.e. $T\equiv0$, can be obtained for $\delta=0$. In practice, however, the transmission never vanishes because of the transport through remote energy levels, which are not taken into account in this model. Evolution of the transmission coefficient profile~(\ref{eq:TraDegen}) with variation of $\delta$ for different ratios of the parameters $\gamma_{s}$ and $\gamma_{a}$ is illustrated in Fig.~\ref{fig10}c-e.

\section{Quantum interference inverters based on $\mathcal{PT}$-symmetric interference transistors}

Consider a quantum analogue of CMOS inverter consisting of two quantum switches, connected between one common output lead and two reference voltage sources with voltages $V_{ref1}$ and $V_{ref2}$, respectively. Input signal $V_{in}$ is applied to the common gate of these switches, which is galvanically isolated from the system. Figure~\ref{fig1} depicts two examples of such quantum interference inverters. For a high-resistance load we can implicitly evaluate the voltage transfer characteristic $V_{out}(V_{in})$ of this inverter and estimate its maximum negative gain, which is achieved at $V_{in}=\frac12(V_{ref1}+V_{ref2})$ due to the symmetry. To do so, the transmission coefficients between the leads $T_{1out}$, $T_{2out}$ and $T_{12}$ should be determined first. Reference voltages $V_{ref1}$ and $V_{ref2}$ (assume $V_{ref1}<V_{ref2}$) are given by some external ideal voltage sources, i.e. we treat them as constants. As the input lead is isolated from the system, the voltage $V_{in}$ influences only transmission coefficients. For high-resistance loads the output voltage $V_{out}$ is derived from the condition $I_{out}=0$, where $I_{out}$ is the total current through the output lead, which is composed of the currents from the first and the second reference voltage leads (with appropriate sign).

Let us consider an inverter composed of two identical quantum switches ($\mathcal{PT}$-symmetric interference transistors). Assuming that resonance width is sufficiently small, we can approximate condition $I_{out}=0$ as follows:
\begin{multline}
\left[f(\varepsilon_{1}-eV_{out})-f(\varepsilon_{1}-eV_{ref1})\right]\times\frac{\left(\gamma_{s}^{2}+\gamma_{a}^{2}\right)\left[\delta_{1}^{2}\left(k_{a}-k_{s}\right)^{2}+4\Gamma^{2}\left(\gamma_{a}^{2}-\gamma_{s}^{2}\right)^{2}\right]}{\delta_{1}^{2}\left(k_{a}-k_{s}\right)^{2}+4\Gamma^{2}\left(\gamma_{s}^{2}+\gamma_{a}^{2}\right)^{2}}\\
=\left[f(\varepsilon_{2}-eV_{ref2})-f(\varepsilon_{2}-eV_{out})\right]\times\frac{\left(\gamma_{s}^{2}+\gamma_{a}^{2}\right)\left[\delta_{2}^{2}\left(k_{a}-k_{s}\right)^{2}+4\Gamma^{2}\left(\gamma_{a}^{2}-\gamma_{s}^{2}\right)^{2}\right]}{\delta_{2}^{2}\left(k_{a}-k_{s}\right)^{2}+4\Gamma^{2}\left(\gamma_{s}^{2}+\gamma_{a}^{2}\right)^{2}}.
\label{eq:ResCondCol}
\end{multline}
Here subscripts $1$ and $2$ correspond to the first and to the second quantum switches. Energies $\varepsilon_{1,2}$ are the energies of degenerate states in the first and in the second system respectively. Assume that they are adjusted to $\varepsilon_{1,2}=eV_{ref1,2}$, i.e. to the biased Fermi level of each reference lead. The applied input voltage influences parameters $\delta_{1,2}$ of the switches. The following model dependence of $\delta_{1,2}$ on the input voltage $V_{in}$ provides a symmetrical transition from the `on'-mode to the `off'-mode of each quantum switch as $V_{in}$ varies in the interval $[V_{ref1},V_{ref2}]$:
\begin{equation}
\delta_{1,2}=\alpha e\left(V_{in}-V_{ref1,2}\right),
\label{eq:TunPar}
\end{equation}
where $0<\alpha<1$ is an electrostatic lever arm  of the input lead (common gate). One can substitute Eq.~(\ref{eq:TunPar}) into Eq.~(\ref{eq:ResCondCol}) and derive the implicit dependence $V_{out}=V_{out}(V_{in})$, which then can be used to get an expression for the maximum gain:
\begin{multline}
G_{max}=G\left(V_{in}=\frac{V_{ref1}+V_{ref2}}{2}\right)\\=\sinh{\frac{e\Delta V}{2kT}}\times\frac{256\alpha^{2}\Gamma^{2}\gamma_{a}^{2}\gamma_{s}^{2}\left(k_{a}-k_{s}\right)^{2}kTe\Delta V}{\left[16\Gamma^{2}\left(\gamma_{s}^{2}-\gamma_{a}^{2}\right)^{2}+\alpha^{2}\left(k_{a}-k_{s}\right)^{2}e^{2}\Delta V^{2}\right]\left[16\Gamma^{2}\left(\gamma_{s}^{2}+\gamma_{a}^{2}\right)^{2}+\alpha^{2}\left(k_{a}-k_{s}\right)^{2}e^{2}\Delta V^{2}\right]}.
\label{eq:GainCol}
\end{multline}
Here $\Delta V=V_{ref2}-V_{ref1}$ is fixed by the external supply voltage. In the saturation regime ($e\Delta V\gg kT$) the maximum value of $G$ grows exponentially with $\Delta V$ due to the factor $\sinh{\frac{e\Delta V}{2kT}}$. For $e\Delta V\lesssim kT$ (in the ohmic regime) it becomes independent of the temperature and we can estimate the minimum difference of the reference voltages (supply voltage) $\Delta V_{crit}$ needed to make the inverter operate, i.e. which provides $G_{max}=1$:
\begin{equation}
\Delta V_{crit}\approx\frac{4\Gamma}{e\alpha\left|k_{a}-k_{s}\right|}\sqrt{4\gamma_{a}^{2}\gamma_{s}^{2}-\gamma_{s}^{4}-\gamma_{s}^{4}-2\gamma_{a}\gamma_{s}\sqrt{5\gamma_{a}^{2}\gamma_{s}^{2}-2\gamma_{s}^{4}-2\gamma_{s}^{4}}}\sim\left|\gamma_{s}-\gamma_{a}\right|\text{ as }\frac{\gamma_{a}}{\gamma_{s}}\rightarrow1.
\label{eq:DVCol}
\end{equation}
From Eq.~(\ref{eq:DVCol}) one can see that $\Delta V_{crit}$ can become infinitesimal as $\gamma_{a}/\gamma_{s}\rightarrow1$. On the other hand, however, $G_{max}$ remains bounded in the ohmic regime even if $\gamma_{a}/\gamma_{s}\rightarrow1$. From the analysis of Eq.~(\ref{eq:GainCol}) one can conclude that for $\Delta V=4\frac{\Gamma}{e\alpha|k_{a}-k_{s}|}\sqrt{|\gamma_{s}^{4}-\gamma_{a}^{4}|}$ the gain $G_{max}$ reaches its maximum: $2(\gamma_{s}/\gamma_{a})^{2}$ for $\gamma_{s}<\gamma_{a}$ or $2(\gamma_{a}/\gamma_{s})^{2}$ for $\gamma_{s}>\gamma_{a}$. Hence, the steepest negative gain of the voltage transfer characteristic is limited to $-2$. Nevertheless, it is suitable for operation of the inverter.

\subsection{Model examples of real molecular structures}
Possible candidates for a physical realization of the proposed quantum switch are molecules with degenerate states, e.g. diradicals,~\cite{bib:DirRev} which are already known for providing transmission antiresonances.~\cite{bib:ZeroDirad} Moreover, linkers can stabilize diradical character of such molecules.~\cite{bib:DirStab} Hence, we can expect that connection of certain contacts to them will not destroy the degeneracy of the states, but rather stabilize it. Diradicals can be classified into two types: disjoint and non-disjoint depending on how their non-bonding orbitals intersect (i.e. whether they have common atomic orbitals or not). It was shown that simple starring procedure can distinguish between these two types.~\cite{bib:Disj-nonDisj1,bib:Disj-nonDisj2} Disjoint diradicals seem to be the most appropriate candidate for our quantum switch. Indeed, applying contacts to atoms comprising different degenerate orbitals means that symmetric and antisymmetric combinations of these orbitals will be connected to the leads by equivalent coupling strength, i.e. parameters $\gamma_{s}$ and $\gamma_{a}$ [introduced in Eq.~(\ref{eq:saCoupl})] in this case can be made equal (at least within the nearest neighbour tight-binding approximation). As was highlighted above, according to Eq.~(\ref{eq:maxTra}) this leads to zero conductance in the `off' state.

Operation principle of quantum interference inverter requires that one switch must be in the `on' state and another in the `off' state. There are two possible ways of dealing with this task. First of all, one can choose two different quantum systems (molecules) to make two quantum switches that is similar to the conventional CMOS, where there are two different types of transistors: n-channel MOS and p-channel MOS. This approach requires a technology of synthesis of two different molecules with strictly given parameters. On the other hand, we can use the same quantum system (molecule) to create both switches, but influence their spectrum in different ways by additional gates. This method needs only one type of molecule to be synthesised, but the introduction of additional gates results in some complication of the conventional technological process. In the following subsections we consider some schematic examples of quantum inverters with the same molecules in both switches. Different energies of the on-site atomic states are assumed to be achieved by a certain configuration of additional gates.

\subsubsection{Model of non-disjoint diradical}
The first example structure we consider is a model of the trimethylenemethane molecule, which is a non-disjoint diradical.~\cite{bib:ZeroDirad} Schematically the quantum inverter structure composed of two such four-atomic (carbon skeleton) molecules is shown in Fig.~\ref{fig1}a. Presented schematic model corresponds to a tight-binding H\"{u}ckel structure of one of the resonance configurations of the trimethylenemethane, which is stabilized as it coincides in symmetry with the leads couplings. Hence, hopping integral $\tau$ is assumed to be greater than $\tau_{1}$ as it corresponds to a higher bond order. The transmission coefficient, phenomenological, and microscopical parameters of such switches are presented in the Supplementary material.

We apply the reference voltages as follows: $V_{ref1}=0$ and $V_{ref2}=V_{0}$ is the supply voltage. The range of the input voltage, thus, is $0\leq V_{in}\leq V_{0}$. Applied input potential changes only some on-site energies of the system (in the shaded region in Fig.~\ref{fig1}a). We take this into account in the following form:
\begin{equation}
\varepsilon_{0}^{1,2}=\varepsilon_{1,2}+\alpha e\left(V_{in}-V_{ref1,2}\right).
\label{eq:VoltInf_3}
\end{equation}
The electrostatic influence of the reference and output leads can also be taken into account in a way similar to~(\ref{eq:VoltInf_3}). It can be shown that this influence only distorts the voltage transfer characteristic and taking it into account is not obligatory to illustrate the operation principles of the quantum interference inverters.

Consider the following example: $\varepsilon_{1}=eV_{ref1}=0$, $\varepsilon_{2}=eV_{ref2}=eV_{0}$, $\alpha=0.5$, $\tau_{1}=1\text{eV}$, $\tau=2\text{eV}$ and $\Gamma=1\text{meV}$. Energies are measured from the Fermi level of the first reference lead. Figure~\ref{fig7}a shows voltage transfer characteristics of the inverter for $V_{0}=5\text{mV}$ and Fig.~\ref{fig7}b -- for $V_{0}=10\text{mV}$ (by dot-dashed lines in both cases). In the latter case the voltage transfer characteristic is obviously better because of higher negative gain achieved.

\subsubsection{Model of disjoint diradical}
\begin{figure}
\centering
\includegraphics[width=0.85\linewidth]{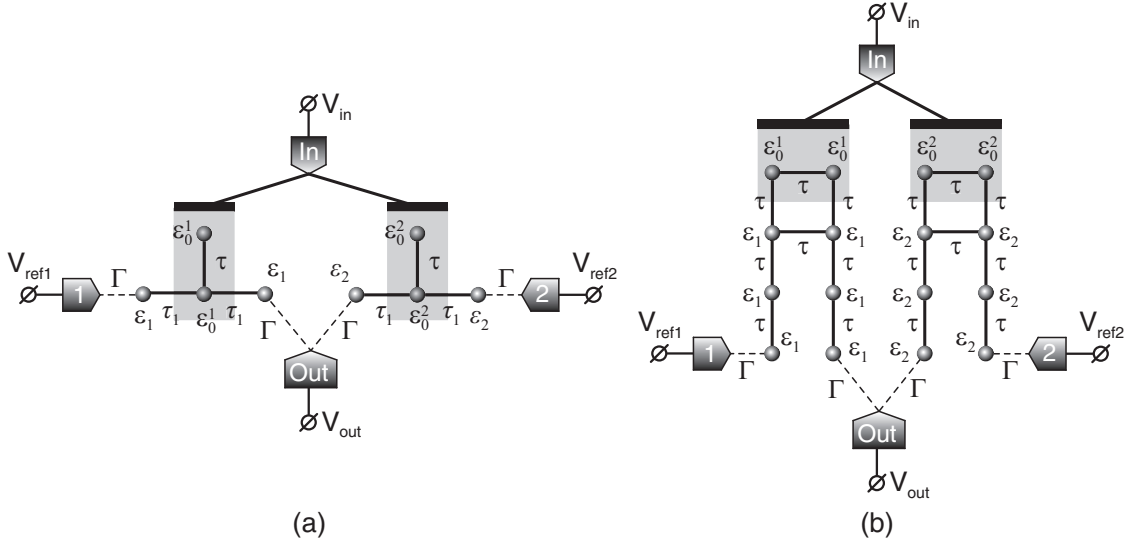}
\caption{Model structures of quantum interference inverter, composed of two quantum switches based on (a) non-disjoint diradicals and (b) disjoint diradicals. Molecules are shown in the form of their H\"{u}ckel theory tight-binding graphs. Shaded regions indicate the sites, which are electrostatically affected by the input gate.}
\label{fig1}
\end{figure}

Another example we consider is a model of the divinylcyclobutadiene molecule, which is a disjoint diradical.~\cite{bib:ZeroDirad} Schematically the quantum inverter structure composed of two such molecules is shown in Fig.~\ref{fig1}b. Presented model corresponds to a simple tight-binding H\"{u}ckel structure of the divinylcyclobutadiene molecule with all bonds treated as equal, providing equal tunnelling matrix elements $\tau$ between p-orbitals of carbon atoms. The transmission coefficient, phenomenological, and microscopical parameters of such switches are presented in the Supplementary material.

\begin{figure}
\centering
\includegraphics[width=0.85\linewidth]{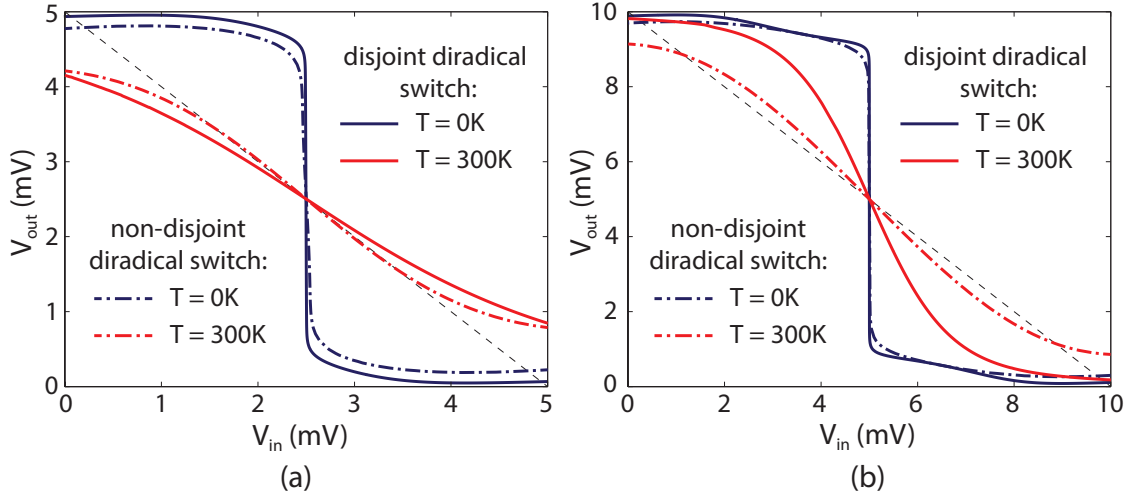}
\caption{Numerically calculated voltage transfer characteristics for the quantum inverter based on $\mathcal{PT}$-symmetric interference transistors for room and zero temperature. Supply voltage is $V_{0}=5\text{mV}$ (a) and $V_{0}=10\text{mV}$ (b). The inverter operates at zero temperature better than at room temperature. Dashed black line shows the $-1$ slope for comparison.}
\label{fig7}
\end{figure}

We assume that the applied input voltage changes only on-site energies in the shaded region in Fig.~\ref{fig1}b, which is taken into account similarly to Eq.~(\ref{eq:VoltInf_3}). Figure~\ref{fig7}a shows voltage transfer characteristics of this inverter for $V_{0}=5\text{mV}$ and Fig.~\ref{fig7}b -- for $V_{0}=10\text{mV}$ (by solid lines in both cases) for the following parameters: $\varepsilon_{1}=eV_{ref1}=0$, $\varepsilon_{2}=eV_{ref2}=eV_{0}$, $\alpha=0.5$, $\tau=1\text{eV}$ and $\Gamma=1\text{meV}$. 

For higher supply voltage transfer characteristic of the inverter based on the disjoint diradical switches (solid lines in Fig~\ref{fig7}b) show higher maximum absolute value of the gain rather than for the inverter based on the non-disjoint diradical switches (dot-dashed lines in Fig~\ref{fig7}b). This is expectable as disjoint diradicals provide $\gamma_{s}/\gamma_{a}=1$ and, thus, the `off'-state current of such switch becomes smaller (it differs from zero only due to the presence of the `background' transmission arising from remote resonance peaks). However, for smaller supply voltage (Fig.~\ref{fig7}a), this `background' component may become high enough to cancel out the key benefit of the disjoint diradical ($\gamma_{s}/\gamma_{a}=1$). Moreover, for lower supply voltages, the range of possible variation of $\delta$ becomes smaller and, due to the non-linearity of energy shifts for disjoint diradicals (i.e. $k_{a,s}$ become functions of $\delta$), the sensitivity to the gate voltage decreases compared to non-disjoint diradicals. In this case the transfer characteristic of a non-disjoint diradical turns out to be more suitable (Fig.~\ref{fig7}a).

\section{Discussion}
We have shown that utilizing the degeneracy of the quantum system spectrum, one can construct the quantum switch operating at infinitesimal supply voltage even at room temperature. Moreover, we propose that a special class of widely studied organic molecules -- diradicals can pretend to make a physical realization of such switches. Thus, this might be a way to dramatically lower the supply voltage, which now cannot be made lower than 0.5-1V\cite{bib:LowVolt1} for the conventional silicon electronic devices, even for promising tunnel field-effect transistors (FET).\cite{bib:LowVolt2} Advance technology of FETs with carbon nanotube (CNT) channel\cite{bib:CNTGraham} also provides a variety of advantages over the bulk Si electronics,\cite{bib:CNTPeng,bib:CNTHan2} but sufficient reduction of the supply voltage is not among them.

However, the question about the noise influence comes up, if we consider low supply voltages and especially sub-$kT/e$ voltages. Strictly speaking, noise in quantum systems is not distinguished into thermal and shot, it is always a superposition of both and it can be described by a closed expression.\cite{bib:ButNoise} Nevertheless, it is illustrative to discuss these contributions independently. Shot noise spectral power is proportional to the current through the system and, thus, it becomes negligible as the voltages and, correspondingly, the currents are scaled down. On the other hand, at finite (room) temperature thermal noise can influence the transport dramatically. Thus, thermal noise is one of the limiting factors of lowering the supply voltage.\cite{bib:LowVolt1}

The mean-square voltage uncertainty is $\Delta V_{therm}=\sqrt{kT/C}$, where $C$ is the capacitance of the load, which is typically the gate capacitance of the next switch. Therefore, using several molecules in parallel in the single switch and, consequently, a bigger gate contact, will increase the capacitance $C$ and lower the noise. But, on the other hand, the greater $C$ is, the worse switching rate $\nu\sim(\tau)^{-1}=(RC)^{-1}$ can be achieved. Here the resistance $R$ can be estimated from the current in the `on' state (see Supplementary material): $R\approx\frac{h}{e^{2}}\times\frac{2kT}{\pi\Gamma}$. Thus, restricting the minimum operating frequency $\nu_{min}$, one can estimate the minimum allowed supply voltage, which we take to be $8$ times the noise voltage uncertainty to provide an error probability at about $10^{-15}$.\cite{bib:LowVolt1} Finally we arrive at the following restriction:
\begin{equation}
V_{0}>\sqrt{2\frac{h}{e^{2}}\times\frac{2(kT)^{2}}{\pi\Gamma}\times\nu_{min}},
\label{eq:NoiseV}
\end{equation}
which for the room temperature and $\Gamma\approx1$meV gives $V_{0}\gtrsim10\text{mV}\times\sqrt{\nu_{min}}$, where $\nu_{min}$ is in GHz. Thus, sub-$kT/e$ supply voltages seems to be possible up to $\nu\approx7$GHz. More detailed analysis of noise impact on operation of quantum interference gates will be presented elsewhere, as well as consideration of technological parameter variation resulting in asymmetry of the inverter structure.

At the moment practical realization of high scale integration of quantum molecular gates is beyond the reach of modern technology. However continuous progress in self-assembling methods and, especially, development of atomic precision lithography could make almost inevitable the implementation of molecules as building blocks of ICs.

\appendix

\section*{Appendix A: Electron occupation symmetry breaking}
\label{ApA}
Real eigenvalues of the auxiliary Hamiltonian define the exact location of perfect transmission resonance and, being $\mathcal{PT}$-symmetric it can experience $\mathcal{PT}$-symmetry breaking, which results in resonance coalescence. This is accompanied by the symmetry breaking of electron occupation in the transmission maximum. The matrix of occupations per unit energy $\hat n$ can be calculated within NEGF formalism:\cite{bib:DattaBook1997}
\begin{equation}
\hat n=\frac{1}{2\pi}\left[f_{L}(\omega)\hat G^{r}\hat \Gamma_{L}\hat G^{a}+f_{R}(\omega)\hat G^{r}\hat \Gamma_{R}\hat G^{a}\right],
\label{eq:OccupDef}
\end{equation}
where $f_{L,R}$ is the Fermi-Dirac distribution function in the left/right lead. Now suppose that symmetric and anti-symmetric states in the site basis are $\ket{s}=(s_{1},s_{2},...,s_{N})^{\top}$ and $\ket{a}=(a_{1},a_{2},...,a_{N})^{\top}$. Thus, neglecting the contribution from distant energy levels, the occupations per unit energy of the $i$-th site (i.e. $(i,i)$ diagonal element of the occupation matrix in the site basis) is following:
\begin{equation}
n_{i}=s_{i}^{2}n_{ss}+a_{i}^{2}n_{aa}+s_{i}a_{i}\left(n_{sa}+n_{as}\right).
\label{eq:ni}
\end{equation}
Here $n_{s,s}$, $n_{s,a}$, $n_{a,s}$, and $n_{a,a}$ are elements of the occupation matrix in the basis of symmetric and anti-symmetric states. If the sites $i$ and $j$ are mapped into each other by the mirror reflection $\sigma_{LR}$ (i.e. $j=\sigma_{LR}(i)$), then corresponding components of the symmetric and anti-symmetric states must be: $s_{i}=s_{j}$ and $a_{i}=-a_{j}$. Therefore, the difference between occupations of this sites is
\begin{equation}
n_{i}-n_{\sigma_{LR}(i)}=2s_{i}a_{i}\left(n_{sa}+n_{as}\right)\propto n_{sa}+n_{as}.
\label{eq:diffOcc}
\end{equation}
This difference appears to be proportional to the sum of non-diagonal elements of the occupation matrix in the symmetric/anti-symmetric states basis. Utilizing Eq.~(\ref{eq:OccupDef}) we can calculate this sum for our system:
\begin{equation}
n_{i}-n_{\sigma_{LR}(i)}\propto n_{sa}+n_{as}=\frac{\Gamma\gamma_{s}\gamma_{s}\left(f_{L}-f_{R}\right)}{\pi\left[4\Gamma^{2}\gamma_{a}^{4}+\left(\omega-\varepsilon-k_{a}\delta\right)^{2}\right]\left[4\Gamma^{2}\gamma_{s}^{4}+\left(\omega-\varepsilon-k_{s}\delta\right)^{2}\right]}\times\left|Q\right|,
\label{eq:diffOccOur}
\end{equation}
where $|Q|$ is given by Eq.~(\ref{eq:Calcs1}). Thus, it is obvious, that at perfect transmission resonances (real zeroes of $Q$) electron occupation is distributed symmetrically (with respect to $\sigma_{LR}$ operation). Whereas, for energies, which correspond to transmission lower than $1$, there is always asymmetric distribution of electrons. Therefore, coalescence of two perfect resonances into one non-perfect is always accompanied by a symmetry breaking of electron distribution, that was shown for linear systems in Ref.~\cite{bib:GorShAop}. 

\bibliographystyle{unsrt}
\bibliography{references}

%
%
%

\end{document}